# Semantic Intelligence in Big Data Applications

Valentina Janev
**Institute Mihajlo Pupin, University of Belgrade, Serbia**
valentina.janev@institutepupin.com

**Summary**
Today, data is growing at a tremendous rate and, according to the International Data Corporation, it is expected to reach 175 zettabytes by 2025. The International Data Corporation also forecasts that more than 150B devices will be connected across the globe by 2025, most of which will be creating data in real-time, while 90 zettabytes of data will be created by the Internet of Things devices. This vast amount of data creates several new opportunities for modern enterprises, especially for analysing the enterprise value chains in a broader sense. To leverage the potential of real data and build smart applications on top of sensory data, IoT-based systems integrate domain knowledge and context-relevant information. Semantic Intelligence is the process of bridging the semantic gap between human and computer comprehension by teaching a machine to think in terms of object-oriented concepts in the same way as a human does. Semantic intelligence technologies are the most important component in developing artificially intelligent knowledge-based systems since they assist machines in contextually and intelligently integrating and processing resources. This Chapter aims at demystifying semantic intelligence in distributed, enterprise and web-based information systems. It also discusses prominent tools that leverage semantics, handle large data at scale and address challenges (e.g. heterogeneity, interoperability, machine learning explainability) in different industrial applications.

## INTRODUCTION

Both researchers and information technology (IT) professionals have to cope with a large number of technologies, frameworks, tools, and standards for the development of enterprise web-based applications. This task has become even more cumbersome with
- the emergence of the **Internet of Things** (IoT, 1999);
- the development of the **Semantic Web** (SW) technologies as a cornerstone for further development of the Web (Berners-Lee 2001);
- the development of **Big Data** solutions (Laney 2001).

Hence, topics such as Smart Data Management (Alvarez 2020), Linked Open Data (Auer et al. 2014), Semantic technologies (Janev and Vraneš 2009) and Smart Analytics have spawned a tremendous amount of attention among scientists, software experts, industry leaders and decision-makers. Table 1 defines a few terms related to data such as open data, big data, linked data and smart data.

Despite the fact that the word Internet of Things ("sensors and actuators embedded in physical objects and connected via wired and wireless networks") is 20 years old, the actual idea of connected devices is older and dates back to 1970s. In the last two decades, with the advancement in information technologies, new approaches have been elaborated and tested for handling the influx of data coming from IoT devices. On one side, the focus in industry has been on manufacturing and producing the right types of hardware to support IoT solutions. On the other, the software industry is concerned with finding solutions that address issues with different aspects (dimensions) of data generated from IoT networks including (1) the *Volume* of data generated by IoT network and the ways of storing the data; (2) the *Velocity* of data and the speed of processing; and (3) the *Variety* of (unstructured) data that is communicated via different protocols and the need for adoption of standards. While these 3 Vs have been continuously used to describe big data, additional dimensions have been added to describe data integrity and quality such as (4) *Veracity* - truthfulness (uncertainty) of data, authenticity, provenance, accountability; (5) *Validity* (correct processing of the data); variability (context of data), (6) *Viscosity* (latency data transmission between the source and destination); (7) *Virality* (speed of the data sent and received from various sources); (8) *Vulnerability* Security and privacy concerns associated with data processing and (9) *Visualization* (interpretation of data and identification of the most relevant information for the users); and (10) *Value* - Usefulness and relevance of the extracted data in making decisions and capacity in turning information into action.

With the rapid development of the Internet of Things, different technologies have emerged to bring the knowledge (Patel et al. 2018) within IoT infrastructures to better meet the purpose of the IoT systems and support critical decision making (Ge et al. 2018; Jain 2021). While the term **Big Data** refers to data sets that have large sizes and complex structures, the term **Big Data Analytics** refers to the strategy of analysing large volumes of data gathered from a wide variety of sources, including different kind of sensors, images/videos/media, social networks, and transaction records. Besides the analytic aspect, Big Data technologies include numerous components, methods and techniques, each employed for a slightly different purpose, for instance, for pre-processing, data cleaning and transformation, data storage and visualization.

**Table 1. Definitions**

| Term | Definition |
| --- | --- |
| Open Data | "The data available for reuse free of charge can be observed as Open Data." (Janev et al. 2018) |
| Big Data | "Big data" is high-volume, velocity, and variety information assets that demand cost-effective, innovative forms of information processing for enhanced insight and decision making." (Laney 2001) |
| | "Big data is high volume, high velocity, and/or high variety information assets that require new forms of processing to enable enhanced decision making, insight discovery and process optimization." (Manyika, 2011) |
| Linked Data | The term Linked Data refers to a set of best practices for publishing structured data on the Web. These principles have been coined by Tim Berners-Lee in the design issue note Linked Data (Berners-Lee 2006). |
| Smart Data | "Simply put, if Big Data is a massive amount of digital information, Smart Data is the part of that information that is actionable and makes sense. It is a concept that developed along with, and thanks to, the development of algorithm-based technologies, such as artificial intelligence and machine learning." (Dallemand 2020) |

Besides the emergence of Big Data, the last decade also witnessed a technology boost for artificial intelligence (AI)-driven technologies. A key prerequisite for realizing the next wave of AI application is to leverage data, which is heterogeneous and distributed among multiple hosts at different locations. Consequently, the fusion of Big Data and IoT technologies and recent advancements in machine learning brought renewed visibility to AI and created opportunities for the development of services for many complex systems in different industries (Tiwari et al. 2018; Mijović et al. 2019). Nowadays, it is generally accepted that AI methods and technologies bring transformative change to societies and industries worldwide.

To reduce the latency, smart sensors (sensor networks) are empowered with **embedded intelligence** that perform pre-processing, reduce the volume and react autonomously. Additionally, in order to put the data in context, standard data models are associated with data processing services, thus facilitating the deployment of sensors and services in different environments.

This Chapter explains the need for semantic standards that improve interoperability in complex systems, introduce the semantic lake concept and demystifying the semantic intelligence in distributed, enterprise and web-based information systems (see Section 2). In order to select an appropriate semantic description, processing model, and architecture solution, data architects and engineers need to become familiar with the analytical problem and the business objectives of the targeted application. Therefore we describe four eras of data analytics and introduces different Big Data tools.

## FORM DATA TO BIG DATA TO SMART DATA PROCESSING

Data-driven technologies such as Big Data and the Internet of Things, in combination with smart infrastructures for management and analytics, are rapidly creating significant opportunities for enhancing industrial productivity and citizen quality of life. As data become increasingly available (from social media, weblogs, IoT sensors, etc.), the challenge of managing (selecting, combining, storing, analyzing) is growing more urgent (Janev 2020). Thus, there is a demands for development of computational methods for the ingestion, management, and analysis of Big Data, as well as for the transformation of this data into knowledge.

From a data analytics point of view, that means that data processing has to be designed taking into consideration the diversity and scalability requirements of the targeted domain. Furthermore, in modern settings, data acquisition occurs in near real-time (e.g. IoT data streams) and the collected and pre-processed data is combined with batch loads by different automated processes. Hence, novel architectures are needed that are is 'flexible enough to support different service levels as well as optimal algorithms and techniques for the different query workloads' (Thusoo 2010).

**Variety of Data Sources.** The development of Big Data-driven pipelines for transforming Big Data into actionable knowledge requires design and implementation of adequate IoT and big data processing architecture, where besides volume and velocity, the variety of available data sources should be considered. The processing and storage of data generated by a variety of sources such as sensors, smart devices, and social media in raw, semi-structured, unstructured, and rich media formats is complicated. Hence, different solutions for distributed storage, cloud computing, and data fusion are needed (Liu 2015). In order to make the data useful for data analysis, companies use different methods to reduce complexity, downsize the data scale (e.g. dimensional reduction, sampling, coding) and pre-process the data (data extraction, data cleaning, data integration, data transformation) (Wang 2017). Data heterogeneity can thus be defined in terms of several dimensions:

- **Structural variety** refers to data representation and indicates multiple data formats and models; for instance, the satellite images format is very different from the format used to store tweets generated on the web;

- **Media variety** refers to the medium in which data gets delivered; for instance, the audio of a speech versus the transcript of the speech may represent the same information in two different media;
- **Semantic variety** refers to the meaning of the units (terms) used to measure or describe the data that are needed to interpret or operate on the data; for instance, a standard unit for measuring electricity is the kilowatt, however the electricity generation capacity of big power plants is measured in multiples of kilowatts, such as megawatts and gigawatts;
- **Availability variations** mean that the data can be accessed continuously; for instance, from traffic cameras, or intermediately, for instance, only when the satellite is over the region of interest.

In order to enable broad data integration, data exchange and interoperability and to ensure that extraction of information and knowledge, standardization at different levels (such as metadata schemata, data representation formats and licensing conditions of open data) is needed. This encompasses all forms of (multilingual) data, including structured and unstructured data, as well as data from a wide range of domains, including geospatial data, statistical data, weather data, public sector information, and research data, to name a few.

**The Need for Semantic Standards.** Michel Bréal, a French philologist, coined the term "semantics" in 1883 to explain how terms may have various meanings for different people depending on their experiences and emotions. In the information processing context, semantics refers to the "meaning and practical use of data" (Woods 1975), i.e., the efficient use of a data object for representing a concept or object. Since 1980, the AI community has promoted the concept of providing general, formalized knowledge of the world to intelligent systems and agents (see also the panel report from 1997 *Data Semantics: what, where and how?*) (Sheth 1997).

Sir Tim Berners-Lee, the Director of the World Wide Web Consortium, presented his vision for the Semantic Web in 2001, describing it as an expansion of the traditional Web and a global distributed architecture where data and services can easily interact. Berners-Lee also introduced the basic (Linked Data) principles for interlinking datasets on the Web via references to common concepts in 2006. The RDF (Resource Description Framework) norm is used to reflect the knowledge that defines the concepts. Parallel to this, increased functionalities and improved robustness of modern RDF stores, as well as wider adoption of standards for representing and querying semantic knowledge, such as RDF(s) and SPARQL, have adopted Linked Data principles and semantic technologies in data and knowledge management tasks. For more information about the recommended Semantic Web technologies, please check the World Wide Web Consortium page at http://www.w3.org/.

Aside from the W3C, there are a few international organizations (associations or consortia) that are important for assessing and standardizing information technologies, such as IEEE-SA (see The Institute of Electrical and Electronics Engineers Standards Association, http://standards.ieee.org/), OASIS (see The Organization for the Advancement of Structured Information Standards, http://www.oasis-open.org/), and a number of others.

**Semantic Integration and Semantic Data Lake Concept.**
The Web, in Tim Berners-vision, is a massive platform-neutral engineering solution that is service-oriented, with service specified by machine-processable metadata, formally defined in terms of messages exchanged between provider and requester agents, rather than the properties of the agents themselves. In the last ten years, businesses have embraced the Tim

Berners-vision and the Linked Data approach and cloud computing infrastructures have enabled the emergence of semantic data lakes.

The following are some of the ways that computer scientists and software providers have tackled the emerging problems in the design of end-to-end data/knowledge processing pipelines:
- In addition to operational database management systems (present on the market since 1970s), different **NoSQL stores** appeared that lack adherence to the time-honored SQL principles of ACID (atomicity, consistency, isolation, durability), see Table 3.
- **Cloud computing** emerged as a paradigm that focuses on sharing data and computations over a scalable network of nodes including end user computers, data centers, and web services (Assunção 2015).
- **Open Data** concept emerged ('data or content that anyone is free to use, reuse and redistribute') as an initiative to enable businesses to use Open Data sources to improve their business models and drive a competitive advantage (see an example of integrating open data in end-to-end processing in modern ecosystem in Figure 1).
- The **Data Lake** concept as a new storage architecture was promoted where raw data can be stored regardless of source, structure and (usually) size. As a result, the data warehousing method (which is built on a repository of centralized, filtered data that has already been processed for a particular purpose) is seen as obsolete, as it causes problems with data integration and adding new data sources.

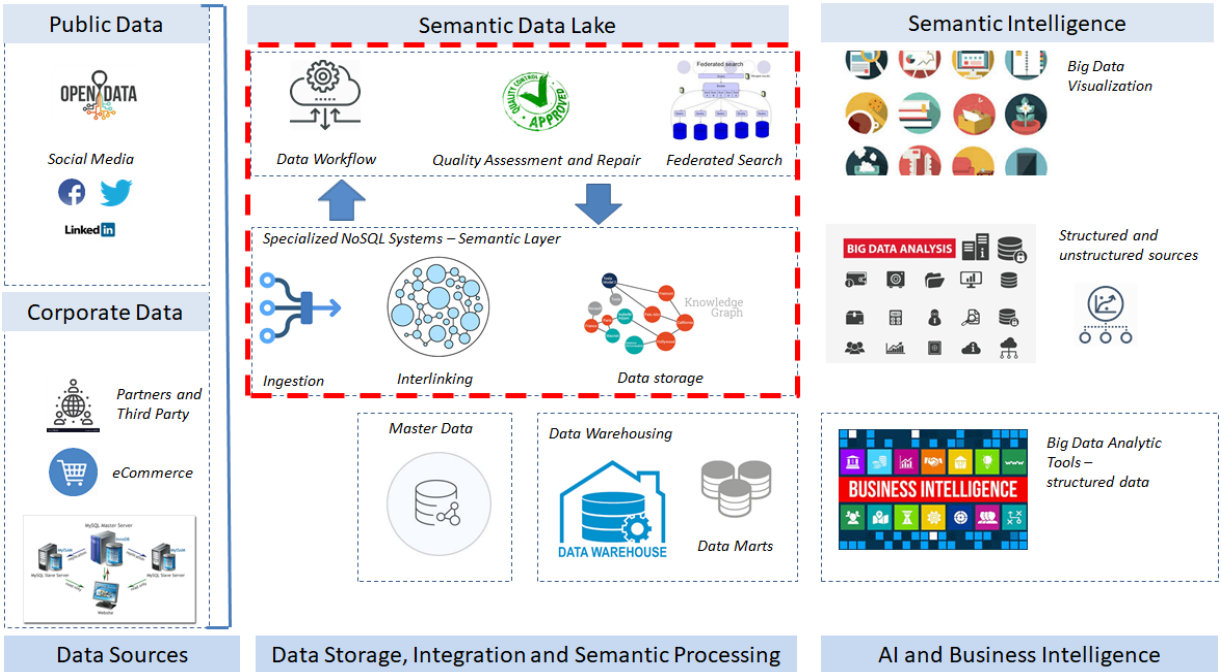

**Figure 1. Modern data ecosystem**

The development of business intelligence services is simple when all data sources collect information based on unified file formats and the data is uploaded to a data warehouse. However, the **biggest challenge that enterprises face is the undefined and unpredictable nature of data appearing in multiple formats.** Additionally, in order to gain competitive advantage over their business rivals, the companies utilize open data resources that are free from restrictions and can be reused, redistributed, and can provide immediate information and insights. **Thus, in a modern data ecosystem, data lakes and data warehouses are both widely used for storing big data.** A *data warehouse* (Kern 2020) is a repository for structured, filtered data that has already been processed for a specific purpose. A *data lake* is a large, raw data repository that stores and manages the company data bearing any format. Moreover, recently, the *semantic data lakes* (Mami 2020) were introduced as an extension of the data lake supplying it with a semantic middleware, which allows the uniform access to original heterogeneous data sources. *Semantic data lakes* integrate knowledge graphs (KGs), a solution that allows to build a common understanding of heterogeneous, distributed data in organizations and value chains and thus to provide smart data for AI applications.

The announcement of the Google Knowledge Graph in 2012 drew a lot of attention to graph representations of general world knowledge. What can be observed in enterprise settings in the last decade is a tendency to collect and encapsulate the metadata in a form of corporate knowledge (or smart data) using semantic technologies while the data is stored or managed via an enterprise knowledge graph (KG). However, many factors have prevented effective large-scale development and implementation of complex knowledge-based scenario because of inability to cope with the rising challenges coming from the Big Data applications; the rigidity of existing database management systems, inability to go beyond the standard requirements of query answering; and the lack of knowledge languages expressive enough to address real-world cases. Despite the challenges, the voluntary created knowledge graphs such as DBpedia (Auer et al. 2007) motivated many big companies, e.g. Google, Facebook, and Amazon, to explore the benefits of using semantic technologies for profit.

## SEMANTICS AND DATA ANALYTICS

Data analytics is a concept that refers to a group of technologies that are focused on data mining and statistical analysis. **Data Analytics** has grown in popularity as a field of study for both practitioners and academics over the last 70 years. The **Analytics 1.0** era started in the 1950s and lasted roughly 50 years. With the advent of Relational Databases in the 1970s and the invention of the Web by Sir Tim Berners-Lee in 1989, the data analytics progressed dramatically as a new software approach and artificial intelligence was developed as a separate scientific discipline.

The Analytics 2.0 era began in the 2000s with the introduction of Web 2.0-based social and crowd-sourcing systems. Although business solutions in the Analytics 1.0 era were focused on relational and multi-dimensional database models, the **Analytics 2.0** era introduced NOSQL and big data database models, which opened up new goals and technological possibilities for analysing large volumes of semi-structured data. Before big data (BBD) and after big data (ABD) are terms used by companies and data scientists to describe these two spans of time (Davenport 2013).

The fusion of internal data with externally sourced data from the internet, different types of sensors, public data projects (such as the human genome project) and captures of audio and video recordings was made possible by a new generation of tools with fast-processing engines and NoSQL stores. The Data Science area (a multifocal field consisting of an intersection of Mathematics and Statistics, Computer Science, and Domain Specific Knowledge) also advanced significantly during this period, delivering scientific methods, exploratory processes, algorithms, and resources that can be used to derive knowledge and insights from data in various forms. The IoT and cloud computing technologies ushered in the **Analytics 3.0** era, allowing for the creation of hybrid technology environments for data storage, real-time analysis, and intelligent customer-oriented services. After the countless possibilities for capitalizing on analytics resources, Analytics 3.0 is also known as *the Era of Impact* or *the*

*Era of Data-enriched offerings* after the endless opportunities for capitalizing on analytics services. For creating value in the data economy, Davenport suggests that the following factors need to be properly addressed:

- combining multiple kinds of information
- adoption of novel information management tools
- introduction of "agile" analytical methods and machine-learning techniques to generate insights at a much faster rate
- embedding analytical and machine learning models into operational and decision processes
- development of skills and processes for data exploration and discovery
- requisite skills and processes to develop prescriptive models that involve large-scale testing and optimization and are a means of embedding analytics into key processes
- leveraging new approaches to decision making and management.

The aim of the **Analytics 4.0** era, also known as *the Era of Consumer-controlled data*, is to give consumers complete or partial control over data. There are various possibilities for automating and augmenting human/computer communications by integrating machine translation, smart reply, chat-bots, and virtual assistants, all of which are associated with the Industry 4.0 trend.

The selection of an appropriate semantic processing model (vocabularies, taxonomies, ontologies that facilitate interoperability) (Mishra and Jain 2020) and analytical solution is a challenging problem and depends on the business issues of the targeted domain, for instance e-commerce, market intelligence, e-government, healthcare, energy efficiency, emergency management, production management, and/or security.

## SEMANTICS AND BUSINESS INTELLIGENCE APPLICATIONS

The topic **Semantic Intelligence** brings together the efforts of the Artificial Intelligence, Machine Learning, and Semantic Web communities. The choice of an effective processing model and analytical approach is a difficult task that is influenced by the business concerns of the targeted domain, for instance risk assessment in banks and the financial sector, predictive maintenance of wind farms, sensing and cognition in production plants, automated response in control rooms, etc. The integration of advanced analytical services with semantic data lakes is a complex and hot research topic (see the eight-step process on Figure 2). Although the aim of semantics is to make data and processes understandable to machines, the goal of semantic intelligence is to make business intelligence solutions accessible and understandable to humans. Natural language processing (NLP) and semantic analysis, for example, are used to understand and address posted questions while incorporating semantic knowledge in human-machine interfaces (digital assistants). In this case NLP methods combine statistical and linguistic methods with **graph-based artificial intelligence.**

**Example.** This example presents the process of creating and publishing a Linked Drug Dataset based on open drug datasets from selected Arabic countries. The Drug Dataset has been integrated in a form of a materialized Knowledge Graph (Lakshen et al. 2020). The overall goal is to allow the business user to retrieve relevant information about drugs from the local company data store and other open source datasets. To that aim an intelligent digital assistant is needed.

The pharmaceutical/drug industry was among the first that validated the Linked data principles and standards recommended by the W3C consortium and used the approach for precise medicine. Table 2 briefly describes the tasks needed for development of a semantic data lake and leveraging AI with the KGs.

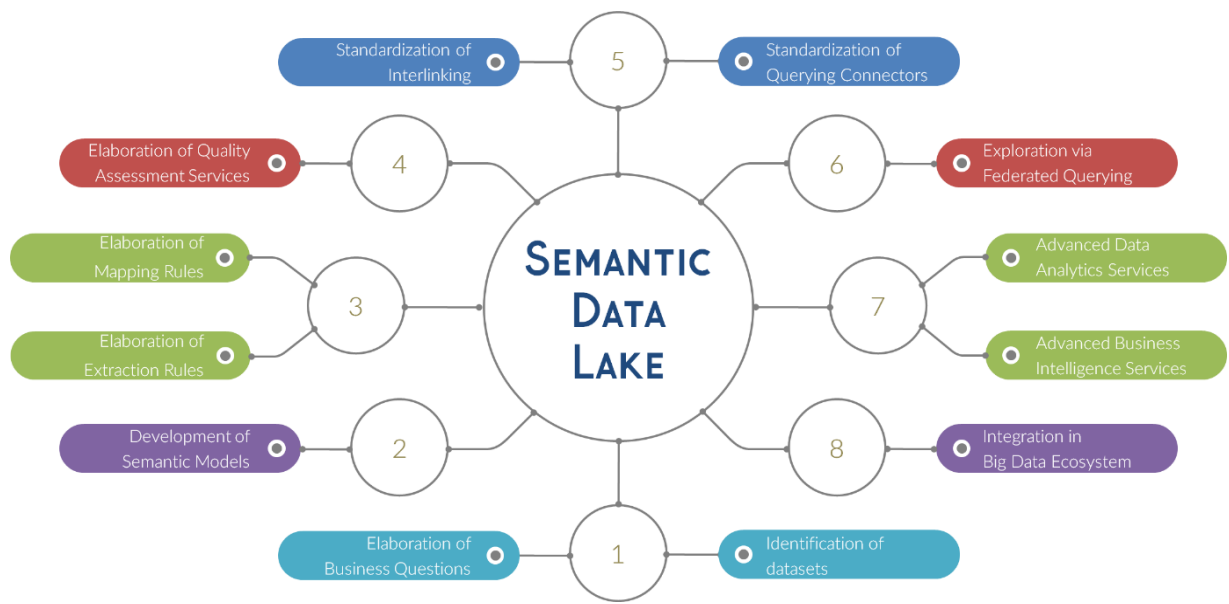

**Figure 2. Semantic Intelligence driven by KGs**

**Table 2. Semantic Intelligence in the Drug Domain (Example)**

| | Step | Description |
|---|---|---|
| 1 | **Identification of datasets** | Data architect first identify the existing company data sources, as well as available open data sources (e.g. DrugBank and DBpedia). |
| | **Elaboration of Business Questions** | The business user specify questions to be answered with a unified access interface to a set of autonomous, distributed, and heterogeneous data sources, as well as with AI-based business intelligence services. |
| 2 | **Development of Semantic Models** | In a case of the drug domain, the drug dataset has properties such as generic drug name, code, active substances, non-proprietary name, strength value, cost per unit, manufacturer, related drug, description, URL, license, etc. Hence, ontology development can leverage re-use of classes and properties from existing ontologies and vocabularies including Schema.org vocabulary[1], DBpedia Ontology[2], UMBEL (Upper Mapping and Binding Exchange Layer)[3], DICOM (Digital Imaging and Communications in Medicine)[4], and DrugBank. |
| 3 | **Elaboration of** | The data administrator runs the extraction process using |

---

[1] https://schema.org/
[2] https://wiki.dbpedia.org/services-resources/ontology
[3] http://umbel.org/
[4] https://www.dicomstandard.org/

|   |                                                      |                                                                                                                                                                                                                                                                                                                                                                                                                                                                                                                      |
|---|------------------------------------------------------|----------------------------------------------------------------------------------------------------------------------------------------------------------------------------------------------------------------------------------------------------------------------------------------------------------------------------------------------------------------------------------------------------------------------------------------------------------------------------------------------------------------------|
|   | **Extraction Rules**                                 | software tools, e.g., OpenRefine (which the authors used), RDF Mapping Language[5], and XLWrap[6] which is a Spreadsheet-to-RDF Wrapper, among others.                                                                                                                                                                                                                                                                                                                                                                 |
|   | **Elaboration of Mapping Rules**                     | For the identified datasets (Excel, XLS data, MySQL store), the data administrator can specify and run mapping rules in order to query the data on-the-fly without data transformation or materialization.                                                                                                                                                                                                                                                                                                             |
| 4 | **Elaboration of Quality Assessment Services**       | The business user / data architect specifies models for describing the quality of semantic (Big Linked) data are needed. Zaveri (2016), for instance, grouped the dimensions into:<br>• *Accessibility*: Availability, licensing, interlinking, security, and performance<br>• *Intrinsic*: Syntactic validity, semantic accuracy, consistency, conciseness, and completeness<br>• *Contextual*: Relevancy, trustworthiness, understandability, and timeliness<br>• *Representational*: Representational conciseness, interoperability, interpretability, and versatility. |
| 5 | **Standardization of Interlinking**                  | Specialized tools are used to help the interlinking and discover links between the source and target datasets. Since the manual mode is tedious, error-prone, and time-consuming, and the fully-automated mode is currently unavailable, the semi-automated mode is preferred and reliable. Link generation application yields links in RDF format using *rdfs:seeAlso* or *owl:sameAs* predicates.                                                                                                                   |
|   | **Standardization of Data Querying Connectors**      | The data administrator specifies connectors as standardized components for interoperability between different solutions. Once the datasets are prepared based on standard vocabularies, the next step is to provide standard querying mechanisms. To that aim, vocabularies such as DCAT and DQV are used to describe the datasets and standardize the access to data. SPARQL is one of the standard querying languages for RDF KGs.                                                                                  |
| 6 | **Exploration via Federated Querying**               | Intelligently searching vast data sets of drug data (patents, scientific publications and clinical trials) data will help, for instance, accelerate the discovery of new drugs and gain insights into which avenues are likely to yield the best results. Federated query processing techniques (Endris et al. 2020) provide a solution to scale up to large volumes of data distributed across multiple data sources. Source details are used to find efficient execution plans that reduce the overall execution time of a query while increasing the completeness of the answers. |
| 7 | **Advanced Data Analytics Services**                 | Drug data aggregated with other biomedical data often display different levels of granularity, that is, a variety of data                                                                                                                                                                                                                                                                                                                                                                                              |

---

[5] https://github.com/RMLio
[6] http://xlwrap.sourceforge.net/

|   |   | dimensionalities, sample sizes, sources and formats. In order to support human decision making, different widgets are needed for visualization and tracing the results of interactive analysis. |
|---|---|---|
|   | **Advanced Business Intelligence Services** | Algorithm-based techniques (machine learning and deep learning algorithms) has already been used in drug discovery, bioinformatics, cheminformatics, etc. What is new in semantic intelligence-based systems is that contextual information from the knowledge graph can be used in machine learning thus improving, for instance, the recommendation and explainability capabilities (Fletcher, 2019; Patel et al. 2020). |
| 8 | **Integration in Big Data Ecosystem** | There are multiple ways of exposing and exploring the KGs-based services to public and other businesses, for instance, using the *Data-as-a-service* or *Software-as-a-service* concept. |

## ROLE OF SEMANTICS IN (BIG) DATA TOOLS

Different keywords are used to name semantic techniques and technologies in the literature and in practice: semantic annotation tools, content indexing and categorization tools; semantic data processing and integration platforms; RDF triple storage systems; semantic web services (Patel and Jain 2019), SOA middleware platforms; semantic annotation tools, content indexing and categorization tools; semantic search and information retrieval technologies; semantic textual similarity methods, linguistic analysis and text mining algorithms, ontology mediated portals; ontological querying/inference engines, rule-based engines; ontology learning methods; ontology reasoners, etc. Davis, Allemang, and Coyne (2004), in their study of the market value of semantic technologies, defined the following four major functions offered by 50 commercial companies in 2004:

- discover, acquire, & create semantic metadata;
- represent, organize, integrate, and inter-operate meanings & resources;
- reason, interpret, infer, & answer using semantics; and
- provision, present, & communicate, and act using semantics.

Janev and Vraneš (2011), based on the analysis of the functionalities of more than 50 SW tools, classified main semantic technology segments into: semantic modelling and creation, semantic annotation, semantic data management and integration, semantic search and retrieval, semantic collaboration including portal technologies, learning and reasoning. Furthermore, in Janev (2020), the author discuss challenges related to Big Data tools and points to a repository of Big Data tools, see https://project-lambda.org/tools-for-experimentation. The tools have been categorized into twelve categories, see also Table 3: Cloud Marketplaces, Hadoop as a Web Service / Platform, Operational Database Management Systems, NoSQL/ Graph databases, Analytics Software / System / Platform, Data Analytics Languages, Optimization Library for Big Data, Library / API for Big Data, ML Library / API for Big Data, Visualization Software / System, and Distributed Messaging System.

From the analysis, we can conclude that it is important to distinguish between Big Data Processing where the size (volume) is one of many important aspects of the data, and Big Data Analytics where semantic processing and use of semantic standards can improve the analysis and produce explainable results.

Table 3. Big Data Tools

| Category | Tools |
|---|---|
| **Cloud marketplaces** | Alibaba Cloud; IBM Cloud; Google Cloud Platform; Oracle Cloud Marketplace; CISCO Marketplace; Microsoft Azure Marketplace; AWS Marketplace |
| **Hadoop as a web service / platform** | HDInsight; IBM InfoSphere BigInsights; MapR; Cloudera CDH; Amazon EMR |
| **Operational database management systems** | IBM (DB2); SAP (SAP HANA); Microsoft (SQL Server); ORACLE (Database) |
| **Nosql/ graph databases** | Hadoop Distributed File System (hdfs); Amazon Neptune neo4j; TigerGraph; Mapr database; OntoText GraphDB; AlegroGraph; Virtuoso; Appache Jena; MarkLogic JanusGraph; OrientDB; Microsoft Azzure Cosmos DB; Apache Hbase; Apache Cassandra; MongoDB |
| **Stream processing engines** | Apache Flume; Apache Apex; Amazon Kinesis Streams; Apache Flink; Apache Samza; Apache Storm; Apache Spark |
| **Analytics software / system / platform** | SAS Analytics Software & solutions; MatLab; H2O.ai; Accord framework; Apache Hadoop; Cloudera data platform; VADALog system; Semantic Analytics Stack (SANSA) |
| **Data analytics languages** | Scala; Julia; SPARQL; SQL; R; Python package index (PyPI); Python |
| **Optimization library for big data** | Facebook ax; Hyperopt; IBM ILOG CPLEX optimization library |
| **Library / API for big data** | TensorFlow serving; MLLIB; BigML; Google Prediction API; Azure machine learning; Amazon machine learning API; IBM Watson programming with Big Data in R |
| **ML library / API for big data** | Caffe.ai; Appache mxnet; Xgboost; PyTorch; Keras; TensorFlow |
| **Visualization software / system** | Oracle Visual Analyzer; Microsoft Power bi; DataWrapper; qlikview; canvas.js; HighCharts; Fusion Chart; d3; Tableau; Google chart |
| **Distributed messaging system** | Apache Kafka |

## SUMMARY

Advances in hardware and software technology, such as the Internet of Things, mobile technologies, data storage and cloud computing, and parallel machine learning algorithms, have allowed the collection, analysis, and storage of large volumes of data from a variety of quantitative and qualitative domain-specific data sources over the last two decades. As presented in this Chapter, interoperable data infrastructure and standardization of data-related technology, including the creation of metadata standards for Big Data management, are needed to simplify and make the Big Data processing more efficient. Semantics play an important role, particularly when it comes to harnessing domain information in the form of Knowledge Graphs. As our analysis showed, in the last decade, especially after the announcement of the Google Knowledge Graph, a large corporations introduced semantic processing technologies to provide scalable and flexible data discovery, analysis, and reporting. The Semantic Data Lake approach have been exploited to allows uniform access to original heterogeneous data, while the semantic standards and principles are used for

- representing (schema and schema-less) data;
- representing metadata (about documentation, provenance, trust, accuracy, and other quality properties);
- modelling data processes and flows, i.e., representing the entire pipeline making data representation shareable and verifiable;
- implementation of standard querying and analysis services.

However, ttransforming Big Data into actionable Big Knowledge demands scalable methods for creating, curating, querying, and analyzing Big Knowledge. Our study on Big Data tools reveals that there are still open issues that impede a prevalence usage of graph based frameworks over more traditional technologies like relational databases and NoSQL stores. For instance, tools are needed for federations of data sources represented using the RDF graph data model for ensuring efficient and effective query processing while enforcing data access and privacy policies. Next, the integration of analytic algorithms over a federation of data sources should be assessed and evaluated. Finally, quality issues that are more likely to be present such as inconsistency and incompleteness should be properly addressed and integrated in the reasoning processes.

Along with the discussion of the emerging Big Data tools on the market (categorized into twelve groups), in this Chapter, we summarized an eight-step approach for utilization of knowledge graphs for semantic intelligence. Hence, we may conclude that there is a broad spectrum of applications in different industries where semantic technologies and machine-learning methods are used for managing actionable knowledge in real-world scenarios.

Once the above mentioned issues are effectively addressed, promising results from semantic intelligence services and applications are expected, for instance, for personalized healthcare, financial portfolio optimization and risk management, and Big Data-driven energy services.

**Acknowledgment**

The research presented in this paper is partly financed by the European Union (H2020 PLATOON, Pr. No: 872592; H2020 LAMBDA, Pr. No: 809965), and partly by the Ministry of Science and Technological Development of the Republic of Serbia and Science Fund of Republic of Serbia (Artemis).